# Stimulated-Raman-Scattering Metrology

M. Lamperti[1,*], L. Rutkowski[2], D. Ronchetti[1,†], D. Gatti[1], R. Gotti[1,‡], G. Cerullo[1], F. Thibault[2], H. Jóźwiak[3], S. Wójtewicz[3], P. Masłowski[3], P. Wcisło[3], D. Polli[1] and M. Marangoni[1]

[1] Dipartimento di Fisica - Politecnico di Milano and IFN-CNR, Via Gaetano Previati 1/C, 23900 Lecco, Italy
[2] Univ Rennes, CNRS, IPR (Institut de Physique de Rennes)-UMR 6251, F-35000 Rennes, France
[3] Institute of Physics, Faculty of Physics, Astronomy and Informatics, Nicolaus Copernicus University, Grudziadzka 5, 87-100 Torun, Poland
* marco.lamperti@uninsubria.it; currently at Dipartimento di Scienza e Alta tecnologia, Università degli studi dell'Insubria, 22100 Como, Italy
† Currently at Department of Physics, Universität Hamburg, Jungiusstrasse 9, 20355 Hamburg, Germany and Max Planck School of Photonics, Friedrich-Schiller University of Jena, Albert-Einstein-Str. 6, 07745 Jena, Germany
‡ Currently at Dipartimento di Ingegneria Industriale e dell'Informazione, Università di Pavia, Via Ferrata 5, 27100 Pavia, Italy

## Abstract
Frequency combs have revolutionized optical frequency metrology, allowing one to determine highly accurate transition frequencies of a wealth of molecular species. Despite a recognized scientific interest, these progresses have only marginally benefited infrared-inactive transitions, due to their inherently weak cross-sections. Here we overcome this limitation by introducing stimulated-Raman-scattering metrology, where a frequency comb is exploited to calibrate the frequency detuning between the pump and Stokes excitation lasers. We apply this approach to molecular hydrogen to test quantum electrodynamics. We measure the transition frequency of the Q(1) fundamental line of $H_2$ around 4155 $cm^{-1}$ with few parts-per-billion uncertainty, which is comparable to the theoretical benchmark of *ab initio* calculations and more than a decade better than the experimental state of the art. Our comb-calibrated stimulated Raman scattering spectrometer extends the toolkit of optical frequency metrology as it can be applied, with simple technical changes, to many other infrared-inactive transitions, over a 50-5000 $cm^{-1}$ range that covers also purely rotational bands.

## Main

The invention of optical frequency combs has enabled, in the past two decades, highly accurate measurements of the energy values of multiple atomic and molecular transitions across a large part of the electromagnetic spectrum [1–3]. One of the areas where precise absolute frequency calibration plays a pivotal role for fundamental physics is the spectroscopic investigation of molecular hydrogen and its isotopologues, whose transition frequencies can be accurately predicted by theory[4]. The comparison between theory and experiments represents a testbed for molecular quantum electrodynamics[5](QED) as well as a compelling approach to explore physics beyond the standard model[6–8] or to determine fundamental quantities such as the nucleon-electron mass ratios[9].

A stumbling block to the accuracy of experimental determinations of transition frequencies is the intrinsic weakness of quadrupole rovibrational transitions, the only ones allowed in homonuclear species such as $H_2$ and $D_2$. Many experiments circumvented this weakness using high-finesse optical cavities providing enhancement of the effective absorption length up to several kilometres[10]. Thanks to the comb referencing of the frequency of the cavity-injected laser, transition frequencies have been measured with uncertainties of few parts per billion (ppb)[11–16], at the same level of *ab initio* calculations[4]. The main limiting factors are the signal-to-noise ratio at low pressures and the nontrivial extrapolation of the line centre to zero pressure because of the complexity of absorption lineshapes[17]. Cavity-enhanced measurements have produced so far accurate determinations only on overtone bands in the near-infrared. Their extension to fundamental rovibrational transitions which fall in the mid-infrared is a complex task, due the decreased quality of lasers, mirrors, detectors and modulators in this region of the spectrum.

An alternative solution for determining the frequency of fundamental transitions is resonantly-enhanced multi-photon ionization (REMPI) applied to a molecular beam: it involves state-selective ionisation of the excited molecule through a pulsed ultraviolet laser and subsequent mass-selective detection of the generated ion. In 2015 Ubachs and co-workers successfully obtained by REMPI an uncertainty of $2 \cdot 10^{-4}$ cm$^{-1}$ on the Q(0), Q(1) and Q(2) lines of the main hydrogen isotopologues[18]. For $D_2$, this benchmark was substantially improved very recently, down to $6 \cdot 10^{-7}$ cm$^{-1}$ (0.2 ppb at a frequency of 3167 cm$^{-1}$) on the S(0) 1-0 transition[19], primarily thanks to an efficient excitation of the molecules in the upper vibrational state and to an accurate reduction of systematic errors related to the fine structure of the transition. The excitation was fostered by a strong static electric field that induces a transition dipole moment in the molecule and thus enhances the absorption of an intense comb-referenced mid-infrared laser. The overall setup remains quite complex and challenging to scale to purely rotational transitions, that would require far-infrared lasers[20].

Here, we tackle the challenge of precision spectroscopy of infrared-inactive vibrational transitions by revitalizing and adding metrological quality to the Stimulated Raman Scattering (SRS) technique[21]. SRS is a third-order nonlinear spectroscopic process that makes use of two laser fields with proper frequency detuning, the so-called pump and Stokes fields, to excite a given vibration. It has been the approach of election in the past century to address transitions of dipole-inactive species like $H_2$ [22,23], but it has never benefited from the enhanced precision afforded by frequency combs. Here we introduce comb-assisted Stimulated-Raman-Scattering Metrology (SRSM) to study with extremely high accuracy fundamental infrared-inactive transitions. The approach has the potential to cover a two-decade-spanning range of frequencies, from 50 to 5000 cm$^{-1}$, by changing the wavelength of the pump laser in the near infrared, leaving all other parts of the apparatus substantially unaltered. We apply the spectrometer to the most studied stretching mode of $H_2$ at ≈ 4155 cm$^{-1}$, corresponding to the Q(1) line of the 1-0 band. We improve the state of the art for $H_2$, obtained through REMPI, by 20 times[18], achieving a combined uncertainty of $1.0 \cdot 10^{-5}$ cm$^{-1}$ (310 kHz) that challenges the theoretical benchmark[4]. This required the implementation of a multi-spectra fitting procedure for the extrapolation to zero pressure of the line centre, with a set of fitting parameters fixed to *ab initio* values inferred from quantum-scattering calculations on $H_2$-$H_2$ collisions[24].

## Results

**Comb-calibrated coherent Raman spectrometer.** SRS requires the use of two detuned lasers (pump field at frequency $\nu_p$ and Stokes field at frequency $\nu_S < \nu_p$) to probe a rovibrational transition at frequency $\nu_p - \nu_S$. An energy transfer from the pump to the Stokes beam occurs when their frequency detuning is resonant with a Raman-active transition, in our case the Q(1) transition of the fundamental rovibrational band of $H_2$ around 4155 cm$^{-1}$. The measured quantity is the so-called Stimulated Raman Loss (SRL), which is the normalized change $\Delta I_p/I_p$ of the pump intensity $I_p$ induced by the Raman interaction. It is proportional, without any limitation given by phase matching, to the interaction length (*L*) and to the Stokes field intensity ($I_S$) [25]:

$$\text{SRL} = \frac{\Delta I_p}{I_p} \propto \text{Im}\{\chi^{(3)}(\nu_p - \nu_S)\} L I_S$$

The resonant third-order nonlinear susceptibility $\chi^{(3)}$ of the medium entails the dependence of the SRL on the pump-Stokes frequency detuning, which reflects in $H_2$ a complex collisional physics. To maximize the SRL signal, we used a multi-pass cell and a high-power Stokes laser.

The spectrometer (depicted in Figure 1) employs a pair of single-frequency lasers for the pump (2 mW at 740 nm) and Stokes (5 W at 1064 nm) fields, with optical frequencies calibrated on a primary frequency standard through an optical frequency comb (see Methods for details). The two beams are superimposed by an active alignment system (see Methods for details) and mode matched to a multi-pass cell filled with $H_2$. The cell can contain gas in a pressure range from 10$^{-3}$ to 5 bar and is equipped with broadband dielectric mirrors that guarantee a total transmission around 50% from 700 to 1100 nm upon 70 reflections inside the cell and a total optical path length of 30 m. Gas temperature and pressure are actively stabilized to user-set values (see

Methods for details). At the cell output the pump is separated from the Stokes and is sent to a photodiode to measure the SRL signal, which is at the $10^{-4}$ level at 1 bar of $H_2$ pressure in our experimental conditions. The signal-to-noise ratio (SNR) of the measurement is brought to the shot-noise limit (corresponding to root-mean square fluctuations of $3.3 \cdot 10^{-8}$ for 1 s measurement time, see Supplementary Section 2 for details) by a modulation transfer technique that exploits high-frequency (9.7 MHz) electro-optical modulation (EOM) of the Stokes intensity and synchronous detection of the SRL with a lock-in amplifier. Spectral acquisitions are carried out by measuring the SRL while repeatedly scanning the Stokes frequency over about 12 GHz around the centre of the transition. The absolute frequency calibration comes from offset-locking the pump laser frequency to the nearest mode of a frequency comb and by real-time tracking of the Stokes frequency through high-speed acquisition and processing of its beat note against the same comb. A single spectral scan takes from 0.5 to 5 s depending on the gas pressure. Spectral points acquired over many subsequent scans with stable thermodynamic conditions are binned to obtain a single averaged spectrum with 1 MHz point spacing.

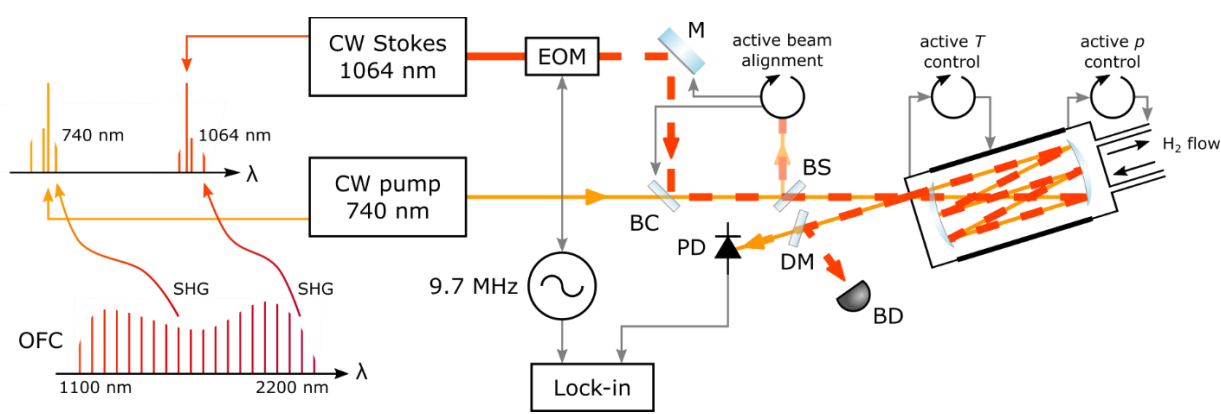

**Figure 1**: **Comb-assisted stimulated Raman scattering.** EOM: electro-optic modulator; M: mirror; BC: beam combiner; BS: beam sampler; DM: dichroic mirror; BD: beam dumper; PD: photodiode; OFC: optical frequency comb; SHG: second harmonic generation. Comb referencing of pump and Stokes lasers is obtained through second harmonic generation of selected spectral portions of an octave-spanning continuum generated by an Er:fibre comb (see Methods for details).

**Spectral measurements.** Spectral acquisitions were made at 9 pressure values in the range 0.05-4 bar and at constant temperature T = 303.1 K. The exploration of such a large pressure range allows one to follow the complex collisional phenomena taking place in the ensemble of gas molecules and enables a robust determination of the transition frequency, as described in the following section and in the Supplementary Material. A typical set of spectra is shown in Figure 2a in an absolute SRL scale. The shape and the width of the line change considerably in the explored pressure range. Qualitatively, the most evident effect is that at high pressure the spectra are narrower and tend to exhibit a Lorentzian lineshape, whereas at lower pressure – see Fig. 2b – the lineshape tends to be broader and Gaussian. This is due to the Dicke effect[26], which is the narrowing of the velocity distribution of the gas molecules and thus of the Doppler profile at increasing collision rates: this effect is prominent in $H_2$ around 2 bar, where it leads to an almost complete extinction of Doppler broadening and to a minimum spectral width of 260 MHz. As a result, the SNR of high-pressure spectra is particularly high even with short measurement times, reaching around 8000 over 10 minutes with a 1 MHz sampling. Another evident qualitative effect is a pronounced shift of the line peak with pressure, which has to be properly accounted for to extrapolate an accurate line centre to zero pressure and compare it with that calculated on the isolated molecule. This calls for a refined modelling of the collisional physics of $H_2$ and for the two-step analysis described below.

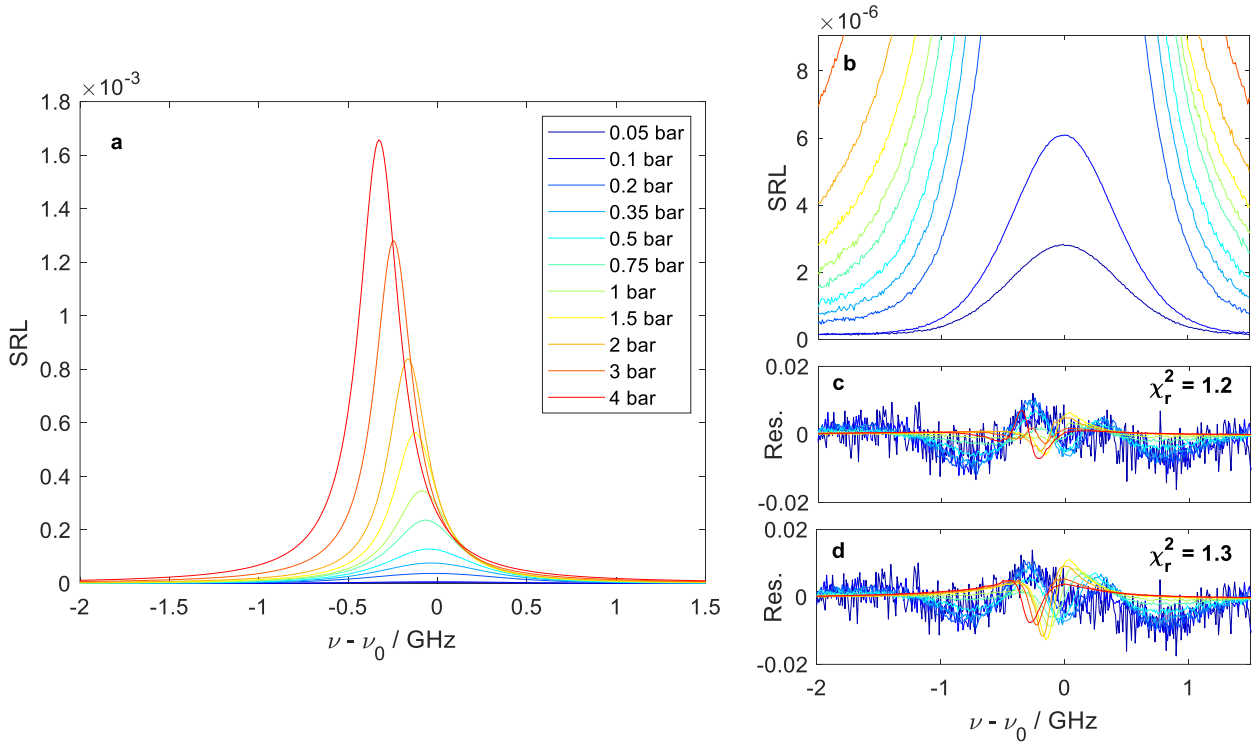

**Figure 2: Stimulated Raman scattering metrology of H$_2$. a,** Set of spectra at different pressures plotted in an absolute SRL scale. **b,** Zoom of panel **a** showing a detail of the spectra acquired at pressures equal to 0.05 and 0.1 bar. **c,** Residuals (data - model) normalized to the corresponding spectral maxima, obtained from a global fitting with 5 free parameters and 3 fixed to *ab initio* values (see Supplementary Section 3 for details). $\chi_r^2$ represents the reduced chi square of the fit. **d,** Residuals as above, obtained from a global fitting with 2 free ($\nu_0$, $\delta_0$) and 6 fixed parameters.

**Multi-pressure global fitting.** As a first step for data analysis we modelled the experimental spectra with the $\beta$-corrected Hartmann-Tran profile ($\beta$HTP), which is a spectral line profile optimized for H$_2$ isotopologues[27]. It accounts for all main collisional phenomena at play, including velocity-changing collisions and speed-dependent effects, while exhibiting an analytical representation that can be efficiently integrated into least-square fitting procedures. To reduce the correlations among the 8 collisional parameters that define the $\beta$HTP profile[24], we fixed a subset of them to *ab initio* values inferred from quantum-scattering calculations of H$_2$-H$_2$ collisions (described in Supplementary Section 3). For every fit we left as free parameters the unperturbed (zero-pressure) transition frequency $\nu_0$, which is the quantity of main interest here, and the linear pressure shift of the line centre $\delta_0$, which is hard to be accurately predicted by theory and can be better constrained by measurements. Figure 2c and d show the residuals of two types of fitting conducted in a global form over the entire multi-pressure dataset, using just 5 (Fig. 2c) or 2 ($\nu_0$, $\delta_0$, Fig. 2d) free parameters, respectively (details given in Supplementary Section 4). In both cases we observe structured residuals at a level of 1% of the line maximum, with extremely similar reduced chi-square values (1.3 and 1.2) and fitted $\delta_0$ parameters (-2.7 and -2.69 cm$^{-1}$). This is an indication of consistency between experimental spectra and lineshape model with *ab initio* collisional calculations. On the other hand, the two fits converge to largely different values of the line centre $\nu_0$, with a discrepancy of almost 2.5 MHz. This is due to the concurrence of two effects, namely the difficulty of $\beta$HTP to fully model the collisional physics of H$_2$ in such a very large pressure range[24] and the inclusion of high-pressure spectra in the fit. While being of key importance for the retrieval of robust collisional parameters, they introduce a strong leveraging effect that amplifies the impact on the line centre of any imperfect assessment of collisional parameters.

**Extrapolation of the zero-pressure line-centre.** The second step in our data analysis was to develop a procedure to extrapolate the line centre overcoming the limitations of the model while reducing the impact of high-pressure spectra. To this purpose, we first investigated how the inferred $\nu_0$ changes upon restricting the fit to experimental data at lower and lower pressures (i.e., progressively discarding high-pressure spectra), down to a value $p_{max}$ = 0.2 bar, to include in the fit at least three pressures. Figure 3 reports $\nu_0$ as a function of $p_{max}$, taking all collisional parameters except $\delta_0$ fixed to *ab initio* values. Three trajectories are retrieved, corresponding to three different fixed values of $\delta_0$, from -2.7 to -2.36 cm$^{-1}$/atm, which define the limits of our confidence interval for this parameter (see Supplementary Section 4 for details on such an interval). They showcase a strong dependence of $\nu_0$ on both $p_{max}$ and $\delta_0$, but at the same time also a clear convergence to one another while reducing $p_{max}$: this suggests an extrapolation of these trajectories to $p_{max}$ = 0 as a viable approach to infer the line centre frequency of the isolated molecule, minimizing the errors introduced by the model and by the uncertainty on $\delta_0$. This extrapolation is shown in Fig. 3b using a quadratic fit over the last four points of the trajectories.

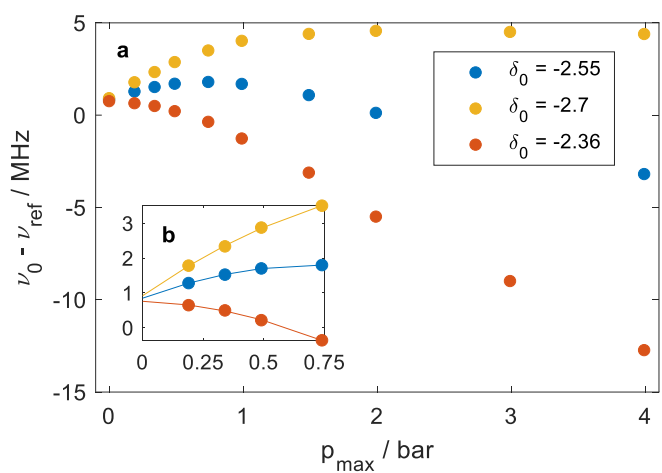

**Figure 3**: **Determination of the frequency of the Q(1) 1-0 line of H$_2$. a,** Inferred $\nu_0$ as a function of the maximum pressure $p_{max}$ considered for the global fitting. The collisional parameters are fixed to *ab initio* parameters except $\delta_0$, which is fixed to the three different values specified in the legend (see text and Supplementary Materials for details), generating the three trajectories shown here. **b,** Zoomed-in view of the low-pressure range together with parabolic extrapolations (lines) to $p_{max}$ = 0. The vertical axis reports the difference between the fitted $\nu_0$ and the *ab initio* value $\nu_{\text{ref}}$ calculated by Komasa et al.[4]

**Uncertainty budget.** The line centre uncertainty due to $\delta_0$ can be quantified with the help of the inset in Fig. 3, which provides a zoomed-in view of the three trajectories and their quadratic extrapolations to $p_{max}$ = 0. The half peak-to-peak distance between the two outer extrapolations amounts to 80 kHz, which was taken as an estimate for this systematic uncertainty term. As a second source of systematic uncertainty, we evaluated the impact on $\nu_0$ of the uncertainty on the *ab initio* parameters kept fixed in the fitting. For every such parameter varied in its confidence interval (see Supplementary Section 4 for details) we calculated trajectories similar to those reported in Fig. 3 and considered the half peak-to-peak distance of the outer extrapolations as the corresponding uncertainty contribution to $\nu_0$, to be summed up in quadrature to the other contributions. This procedure leads to an additional term of systematic uncertainty of 220 kHz. To quantify the statistical uncertainty, we applied a bootstrap procedure taking advantage of the many spectral datasets acquired, running many fits on randomly assembled spectral datasets obtained by combinations of spectra acquired in different days over several months. This allows to account not only for white noise sources on the vertical and horizontal scales of the measurement, but also for contributions due to long-term drifts in the spectrometer or to uneven thermodynamic conditions. Differently from the systematic uncertainties above, we used here the root-mean square deviation of the different fits as an estimator for a final value of 200 kHz. The combined uncertainty budget for the extrapolated transition frequency, obtained by quadrature sum of the three contributions above, amounts to 310 kHz (10$^{-5}$ cm$^{-1}$), thus half that of the theoretical benchmark for the transition[4] and 20 times better than the experimental benchmark[18].

## Discussion

The measured transition frequency is 4155.253790(10) cm$^{-1}$. It is compared in Table 1 with the best experimental[18] and theoretical[4] determinations reported so far. For both cases the agreement with our determination is within 1σ, using the combined uncertainty of the two compared frequencies as an estimator for σ.

**Table 1.** Comparison between the frequency of the Q(1) 1 - 0 transition obtained in this work with the best available experimental[18] and theoretical[4] data. The deviation Δ between the frequencies is calculated by subtracting others' data from our value. The combined uncertainty σ is calculated as the quadrature sum of the uncertainties of the two measurements. Note: it is a coincidence that deviations equal uncertainties.

|                | This work          | Experiment[18]    | Theory[4]          |
|----------------|--------------------|--------------------|--------------------|
| $\nu_0$ / cm$^{-1}$ | 4155.253790(10)    | 4155.25400(21)     | 4155.253762(26)    |
| $\nu_0$ / MHz  | 124571374.73(31)   | 124571381.0(63)    | 124571373.89(78)   |
| Δ / MHz        | /                  | -6.3               | 0.84               |
| σ / MHz        | /                  | 6.3                | 0.84               |

To put this result in a broader context, we report in Figure 4 the deviations between theory and experiments for all transition frequencies accurately measured so far on neutral molecular hydrogen and its isotopologues. Interestingly, experimental determinations are in excess of calculated values in all cases, independently of the isotopologue, of the rovibrational branch, of the experimental approach and of whether the transition is dipole or quadrupole allowed, with discrepancies from 0.4 up to 2.3σ. Focusing on quadrupole lines, by far the most accurate experimental determinations were obtained on $D_2$, with an uncertainty down to 161 kHz for lines of the 2-0 band using cavity-ring-down spectroscopy[11], and of 17 kHz for lines of the 1-0 band on a molecular beam[19], in both cases well below the theoretical uncertainty. For $H_2$, the centre frequency here reported is the most accurate obtained so far and the only value exhibiting an accuracy comparable with theory, thus making it of significance for QED tests. The discrepancy with theory is of the same sign and of the same order of magnitude (1σ) encountered in several measurements on $D_2$, and is dominated by the theoretical uncertainty. Overall, experimental accuracy has grown up at a faster pace than theory in recent years, reaching and even surpassing theory in some cases, yet not at point to challenge our current QED modelling of molecules. This challenge requires further efforts, over both axes.

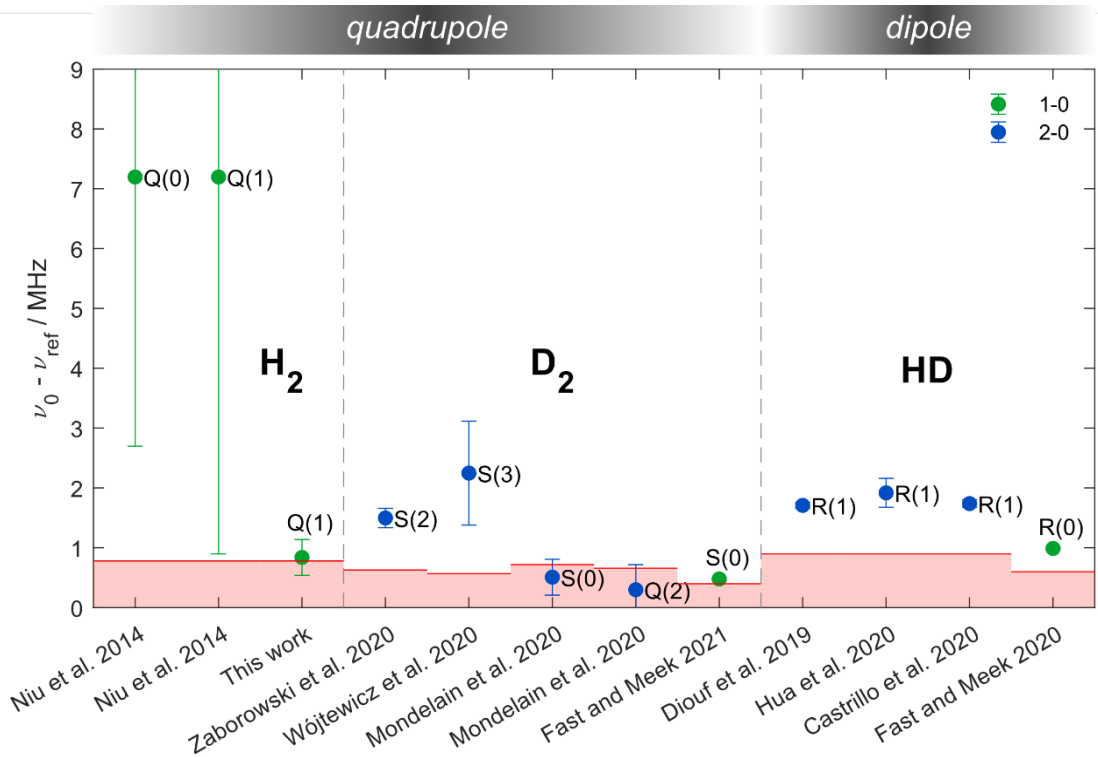

**Figure 4**: **Comparison of experiment and theory for various rovibrational transitions of H$_2$ isotopologues.** Data points represent the discrepancy on the transition frequency (experiment – theory), with error bars given by the experimental uncertainty and the height of the red shaded area at the bottom representing the theoretical uncertainty. Line names are reported next to data points, while the rovibrational band is identified through the point colour (see legend).

In conclusion, our work brings SRSM in the arena of available techniques for tests of fundamental physics in molecular systems. For the Q(1) 1-0 line of H$_2$ we demonstrated an improvement by more than one order of magnitude with respect to the most accurate experimental determination, and surpassed the uncertainty of state-of-the-art theoretical calculations by a factor of 2. The strength of our setup lies in the ability to target a great variety of rovibrational transitions, in the mid or far infrared, by simple tuning of a near-infrared cw laser. This metrological tool enables probing other fundamental rovibrational transitions of H$_2$ and its isotopologues. It also opens up for addressing Raman-active transitions of heavier molecules, as well as extending comb-calibration to so far unaddressed purely rotational bands.

## Methods

**Absolute frequency calibration of Raman spectra.** The frequency detuning between the pump and Stokes laser fields is calibrated against a frequency comb generated by an amplified Er:fibre femtosecond oscillator emitting pulses at a repetition frequency $f_{\text{rep}} = 100$ MHz, stabilized on a GPS-disciplined Rb clock. The experimental configuration is described in detail in Supplementary Section 1. Briefly, we first broaden the comb inside a highly nonlinear fibre, then frequency double (through second harmonic generation in periodically poled nonlinear crystals) spectral regions around 1480 nm and 2128 nm to obtain narrow frequency combs around 740 nm and 1064 nm. These are mixed with the pump and Stokes beams, respectively, to obtain beat notes between the comb and each cw laser. If we write the frequency of the $n$-th tooth of the broadened comb as $\nu_n = n f_{\text{rep}} + f_{\text{ceo}}$, the frequencies of the pump and Stokes lasers can be expressed as $\nu_{\text{p/s}} = n_{\text{p/s}} f_{\text{rep}} + 2 f_{\text{ceo}} \pm f_{\text{beat,p/s}}$, where the carrier-envelope frequency $f_{\text{ceo}}$ is doubled by second harmonic generation and the beat note $f_{\text{beat,p/s}}$ is added or subtracted depending on the relative position of the pump/Stokes frequencies with respect to their closest comb tooth of index $n_{\text{p/s}}$. The detuning between the two cw lasers can thus be unambiguously determined from the relationship $\nu_{\text{p}} - \nu_{\text{s}} =$

$(n_\text{p} - n_\text{s})f_\text{rep} \pm (f_\text{beat,p} - f_\text{beat,s})$ once the pump and Stokes beat notes are measured and the correct comb-index detuning $\Delta n = (n_\text{p} - n_\text{s})$ is identified, with no contribution from $f_\text{ceo}$. In our case the pump beat note is stabilized at 10 MHz while that of the Stokes beat note is measured in real time while scanning the frequency, as described in the next paragraph. The integer number $\Delta n$ is determined by minimizing the distance between the experimental line centre and its reference value from literature.

**Spectra acquisition and processing.** During spectral acquisitions the pump frequency is kept constant while the Stokes frequency is repeatedly scanned over about 12 GHz with a rate from 1 Hz at high pressure to 0.1 Hz at low pressure[28]. To ensure the proper calibration of spectra, we use a single digital acquisition board to synchronously measure the SRL signal and the beat note of the Stokes field. The Stokes beat note is digitized at 14 bit and 100 MSamples s$^{-1}$ and its frequency calculated in real-time by FFT applied at every 1024 samples via a field-programmable gate array available for on-board processing (FPGA, PXIe-7961 FPGA board and NI-5781 add-on, National Instruments). The demodulated SRL signal is simultaneously acquired and averaged over the same 1024 samples. The absolute frequency is reconstructed in post-processing by unwrapping the measured beat note barycentre[28,29]. Spectral acquisition times typically vary from 10 to 30 min depending on the pressure, but longer times could be used to further enhance the SNR. The spectral points pertaining to subsequent scans are binned to produce the final averaged spectra with frequency spacing of 1 MHz. Fitted spectra and residuals are further binned with bin width of 10 MHz to help identification of non-flat residuals (see example in Figure 2).

**Active alignment of beams.** Any angular misalignment between pump and Stokes beams during the Raman interaction was found to originate a shift of the measured frequency detuning[28]. Manual alignment is accurate to about 300 µrad in our setup, corresponding to fluctuations of the measured centre frequency by about 1-2 MHz from measurement to measurement. To ensure optimal superposition of the laser beams, we built an active alignment system based on the imaging of the near and far fields of both beams on a camera to track any displacement and correct it in a servo loop that acts on piezo-tilt mirror holders. To this purpose, the pump and Stokes beams are sampled right after the beam combiner (BC in Figure 2), then sent through a beam splitter to generate two replicas. The first replica impinges on the sensor of a colour CMOS camera after passing through a lens that provides imaging of the BC plane onto the sensor; we refer to this plane as the near field (NF). The second replica impinges on the sensor after a total propagation distance equal to that from the beam combiner to the centre of the cell, that we refer to as the far field (FF). The differential sensitivity of the colour pixels of the camera at pump and Stokes wavelengths allow disentangling the two colours and control their relative displacements in both the NF and FF. This takes place in real time thanks to a LabView software that, upon calculation of the pump and Stokes centroids from the fitting of their intensity profiles, optimizes the overlap of the Stokes beam onto the pump beam through a 4-channels PID controller followed by 4 PZT tip-tilt mirror adjusters. The system provides an angular RMS stability of 6 µrad (corresponding to RMS fluctuations of the measured frequency detuning by 32 kHz) over few hours.

**Active stabilization of thermodynamic gas parameters.** Uniform and constant temperature of the gas inside the multipass cell is obtained by encasing the cell inside a box made with thick Styrofoam. A pair of rubber heater stripes is placed onto the cell central cylinder, and two fans circulate the air inside the Styrofoam box to ensure temperature homogeneity better than 100 mK on the outer surface of the cell. The temperature is measured through a calibrated temperature sensor (a Pt100 probe paired to a 6 ½ digit multimeter) with an accuracy of 50 mK, whose output is used by a software PID controller (implemented in LabView) to regulate the current passing through the heater stripes and stabilize the temperature to 303.1 K. To compensate for small leaks of the cell, a constant flow of about 10$^{-2}$ L/min is established in the cell using two flow controllers, the first one installed between the H$_2$ cylinder and the cell, the second one between the cell and the vacuum pump. The pressure inside the cell is measured via a calibrated pressure sensor with relative uncertainty better than 10$^{-3}$. Through a software PID control loop, the output flow is regulated to maintain a constant pressure inside the cell.

## Data availability
Averaged spectra and MATLAB code used for fitting will be made available through Zenodo.

## References

1. Holzwarth, R. *et al.* Optical Frequency Synthesizer for Precision Spectroscopy. *Phys. Rev. Lett.* **85**, 2264–2267 (2000).
2. Diddams, S. A. *et al.* Direct link between microwave and optical frequencies with a 300 THz femtosecond laser comb. *Phys. Rev. Lett.* **84**, 5102–5105 (2000).
3. Fortier, T. & Baumann, E. 20 Years of Developments in Optical Frequency Comb Technology and Applications. *Commun. Phys.* **2**, 1–16 (2019).
4. Komasa, J., Puchalski, M., Czachorowski, P., Łach, G. & Pachucki, K. Rovibrational energy levels of the hydrogen molecule through nonadiabatic perturbation theory. *Phys. Rev. A* **100**, (2019).
5. Salumbides, E. J., Dickenson, G. D., Ivanov, T. I. & Ubachs, W. QED effects in molecules: Test on rotational quantum states of $H_2$. *Phys. Rev. Lett.* **107**, 2–5 (2011).
6. Salumbides, E. J. *et al.* Bounds on fifth forces from precision measurements on molecules. *Phys. Rev. - Part. Fields Gravit. Cosmol.* **87**, 1–8 (2013).
7. Ubachs, W., Koelemeij, J. C. J., Eikema, K. S. E. & Salumbides, E. J. Physics beyond the Standard Model from hydrogen spectroscopy. *J. Mol. Spectrosc.* **320**, 1–12 (2016).
8. Biesheuvel, J. *et al.* Probing QED and fundamental constants through laser spectroscopy of vibrational transitions in HD+. *Nat. Commun.* **7**, 10385 (2016).
9. Patra, S. *et al.* Proton-electron mass ratio from laser spectroscopy of HD+ at the part-per-trillion level. *Science* **369**, 1238–1241 (2020).
10. Romanini, D., Ventrillard, I., Méjean, G., Morville, J. & Kerstel, E. Introduction to Cavity Enhanced Absorption Spectroscopy. in *Cavity-Enhanced Spectroscopy and Sensing* (eds. Gagliardi, G. & Loock, H.-P.) 1–60 (Springer, 2014). doi:10.1007/978-3-642-40003-2_1.
11. Zaborowski, M. *et al.* Ultrahigh finesse cavity-enhanced spectroscopy for accurate tests of quantum electrodynamics for molecules. *Opt. Lett.* **45**, 1603 (2020).
12. Wójtewicz, S. *et al.* Accurate deuterium spectroscopy and comparison with ab initio calculations. *Phys. Rev. A* **101**, 052504 (2020).
13. Mondelain, D., Kassi, S. & Campargue, A. Transition frequencies in the (2-0) band of D2 with MHz accuracy. *J. Quant. Spectrosc. Radiat. Transf.* **253**, 107020 (2020).
14. Diouf, M. L., Cozijn, F. M. J., Darquié, B., Salumbides, E. J. & Ubachs, W. Lamb-dips and Lamb-peaks in the saturation spectrum of HD. *Opt. Lett.* **44**, 4733 (2019).
15. Hua, T.-P., Sun, Y. R. & Hu, S.-M. Dispersion-like lineshape observed in cavity-enhanced saturation spectroscopy of HD at 14 μm. *Opt. Lett.* **45**, 4863 (2020).
16. Castrillo, A., Fasci, E. & Gianfrani, L. Doppler-limited precision spectroscopy of HD at 1.4 μm: An improved determination of the R (1) center frequency. *Phys. Rev. A* **103**, 22828 (2021).
17. Wcisło, P., Gordon, I. E., Cheng, C. F., Hu, S. M. & Ciuryło, R. Collision-induced line-shape effects limiting the accuracy in Doppler-limited spectroscopy of H2. *Phys. Rev. A* **93**, 022501 (2016).
18. Niu, M. L., Salumbides, E. J., Dickenson, G. D., Eikema, K. S. E. & Ubachs, W. Precision spectroscopy of the $X^1 \Sigma_g^+$, v = 0 → 1 (J = 0 - 2) rovibrational splittings in $H_2$, HD and D2. *J. Mol. Spectrosc.* **300**, 44–54 (2014).
19. Fast, A. & Meek, S. A. Precise measurement of the $D_2$ $S_1$ (0) vibrational transition frequency. *Mol. Phys.* e1999520 (2021) doi:10/gnrkqd.

20. Fast, A. & Meek, S. A. Sub-ppb Measurement of a Fundamental Band Rovibrational Transition in HD. *Phys. Rev. Lett.* **125**, 023001 (2020).
21. Eesley, G. L. Coherent raman spectroscopy. *J. Quant. Spectrosc. Radiat. Transf.* **22**, 507–576 (1979).
22. Rahn, L. A. & Rosasco, G. J. Measurement of the density shift of the H 2 *Q* (0–5) transitions from 295 to 1000 K. *Phys. Rev. A* **41**, 3698–3706 (1990).
23. Forsman, J. W., Sinclair, P. M., Duggan, P., Drummond, J. R. & May, A. D. A high-resolution Raman gain spectrometer for spectral lineshape studies. *Can. J. Phys.* **69**, 558–563 (1991).
24. Wcisło, P. *et al.* Accurate deuterium spectroscopy for fundamental studies. *J. Quant. Spectrosc. Radiat. Transf.* **213**, 41–51 (2018).
25. Polli, D., Kumar, V., Valensise, C. M., Marangoni, M. & Cerullo, G. Broadband Coherent Raman Scattering Microscopy. *Laser Photonics Rev.* **12**, 1800020 (2018).
26. Dicke, R. H. The effect of collisions upon the doppler width of spectral lines. *Phys. Rev.* **89**, 472–473 (1953).
27. Konefał, M. *et al.* Analytical-function correction to the Hartmann–Tran profile for more reliable representation of the Dicke-narrowed molecular spectra. *J. Quant. Spectrosc. Radiat. Transf.* **242**, (2020).
28. Lamperti, M., Rutkowski, L., Gatti, D., Gotti, R. & Marangoni, M. Manuscript in preparation.
29. Lamperti, M. *et al.* Optical frequency metrology in the bending modes region. *Commun. Phys.* **3**, (2020).

## Acknowledgements
This work was supported by the Italian FARE-MIUR project CH2ROME (Grant no. R164WYYR8N)

## Author contributions
M.M. conceived the experiment. M.L. and L.R. developed the spectrometer. M.L. and D.R. performed the measurements. D.G. developed the real-time acquisition system for the Stokes beat note and the SRS signal. M.L. conceived the line centre extrapolation strategy to zero pressure. S.W., L.R., M.L. performed the multi-pressure HTP fitting. P.W., F.T. and H.J. performed quantum-scattering calculations of collisional parameters. D.P., G.C, P.M., P.W., D.G., R.G. supervised the study. M.L. and M.M. wrote the manuscript with input from all authors.

## Competing interests
The authors declare no competing interests.